\documentclass[prb,twocolumn,showpacs,floatfix]{revtex4-1}

\usepackage{graphicx}
\usepackage{amssymb,amsmath}
\usepackage[colorlinks,citecolor=blue,urlcolor=blue]{hyperref}
\usepackage{dcolumn}

\begin{document}

\title{Interplay between the edge-state magnetism and long-range Coulomb interaction 
in zigzag graphene nanoribbons: quantum Monte Carlo study}

\author{Marcin Raczkowski}
\author{Fakher F. Assaad}
\affiliation{Institut f\"ur Theoretische Physik und Astrophysik,
             Universit\"at W\"urzburg, Am Hubland, D-97074 W\"urzburg, Germany}

\date{\today}

\begin{abstract} 
We  perform projective quantum Monte Carlo simulations of zigzag graphene nanoribbons within a realistic model with long-range Coulomb 
interactions. Increasing the relative strength of nonlocal interactions with respect to the on-site repulsion does not generate a phase transition 
but has a number of nontrivial effects. At the single-particle level we observe  a marked enhancement of the Fermi  velocity  at the Dirac points. 
At the two-particle level,   spin- and charge-density-wave  fluctuations compete. As a consequence, the edge magnetic moment is reduced but the edge 
dispersion relation increases in the sense that the single-particle gap at momentum $q=\pi/|{\pmb a}_1|$  grows.  
We attribute this to nonlocal charge fluctuations  which assist the spin fluctuations to generate the aforementioned gap. 
In contrast, the net result of the interaction-induced renormalization of different energy scales is a constant spin-wave velocity  of the edge modes. 
However, since the particle-hole continuum is shifted to higher energies---due to the renormalization  of the Fermi velocity---Landau damping is reduced. 
As a result, a roughly linear spin-wave-like mode at the edge spreads out through a larger part of the Brillouin zone.

\end{abstract}

\maketitle

\section{Introduction}
\label{sec:intro}

Quasi-one-dimensional graphene nanoribbons terminated by zigzag edges feature localized edge states 
at the Fermi energy.~\cite{Nakada96,Wakabayashi99,Brey06}
Mean-field studies of the Hubbard model, with an on-site effective interaction only,~\cite{Wakabayashi96,Fernandez08,Jung09} 
as well as density-functional theory calculations,~\cite{Son06,Pisani07,Yazyev08} predict spontaneously 
induced spin polarizations at the zigzag edges. The resulting finite dispersion of the low-energy electronic 
states is usually considered in the scanning tunneling spectroscopy as a signature of edge-state magnetism,~\cite{Tao11}
despite merely a weak paramagnetic signal measured in the bulk graphene samples.~\cite{Sepioni10} 
The distinct low-energy electronic structure and the consequent transport properties of zigzag nanoribbons 
have been the subject of great interest.~\cite{Wakabayashi09,Yazyev10}
Moreover, progress in atomic engineering makes it possible nowadays to build nanoribbons with atomically precise zigzag 
edges~\cite{Fasel16} and to verify an early promising proposal of their applicability in nanoelectronics.~\cite{Son06a,Rycerz07}

The edge magnetism of zigzag graphene nanoribbons has also been extensively studied beyond the mean-field level 
using the density matrix renormalization group,~\cite{Hikihara03,Legeza16} exact diagonalization,~\cite{Luitz11}
and quantum Monte Carlo (QMC) simulations.~\cite{Feldner11}
A substantial ferromagnetic (FM)  spin correlation length along the zigzag edge found in the latter study already for 
ribbons of moderate widths, i.e., with four zigzag chains, justifies the picture obtained within 
the mean-field type approximations.~\cite{Feldner11}  In addition,  an effective antiferromagnetic (AF) coupling between 
the opposite edges of the nanoribbon  guarantees the spin-singlet nature of the ground state in accordance with Lieb's theorem 
for bipartite lattices with a half-filled band.~\cite{Lieb89} Thus, zigzag graphene  nanoribbons constitute an interesting 
many-body system with a novel type of boundary-critical phenomena.~\cite{Affleck12}

However, in the semi-metallic state, as realized in graphene, the Coulomb interaction is only partially screened and remains long-range.~\cite{Kotov12}  
As shown within the leading-order perturbative renormalization group analysis~\cite{Gonzalez94,Gonzalez99} 
and confirmed by more elaborated calculations,~\cite{Barnes14,Hofmann14,Bauer15,Sharma16} this results in 
a marginal Fermi liquid behavior and a logarithmic renormalization of the electron velocity at the Dirac points.
The predicted renormalization of the Fermi velocity has been observed in suspended graphene using cyclotron 
resonance~\cite{Elias11} and has triggered intense theoretical studies of the effects driven by nonlocal Coulomb 
interactions in graphene and related materials.~\cite{Semenoff12,Wu14,Smekal14,Golor15,Adam15,Faye15,Classen16,Scherer16,Katanin16,Honerkamp17,Prokofev17}  
A detailed analysis of this fundamental problem that includes the effect of both the short- and long-range parts of the Coulomb repulsion 
on the velocity renormalization can be found in Ref.~\onlinecite{Tang17}.

Given the long-range character of the Coulomb interaction in graphene, it stands to reason that 
there is a need to revisit the stability and dynamical signatures of spin-polarized edge states of zigzag graphene
nanoribbons in the presence of nonlocal interactions.
Only very recently, a step towards a realistic description of the nanoribbons beyond the mean-field treatment 
of Coulomb interactions~\cite{Yamashiro03,Wunsch08,Jung11} was taken by Shi and Affleck.~\cite{Affleck17} 
By neglecting the dynamics of low-lying bulk states, they derived an effective Hamiltonian with nonlocal interactions 
projected onto the Hilbert space of edge states. Within this framework, they were able to provide
a sufficient condition under which the edge magnetism is stable against unscreened Coulomb interactions.

The aim of this paper is to validate those conclusions by performing lattice simulations of the full Hamiltonian 
with the unscreened $\sim 1/r$ tail of the electron-electron repulsion. 
To this end, we carry out systematic studies of zigzag graphene nanoribbons with up to $N=480$ lattice sites  
using the unbiased QMC technique capable of dealing with long-range density-density interactions.~\cite{Hohenadler14} 
Our results lend further support for the robustness of the edge magnetism.  We also find that nonlocal interactions 
play an important role in the renormalization of the dispersion of edge states.
Since the screening length in graphene can be efficiently engineered by substrate modification,~\cite{Hwang12}  
deposition of a zigzag nanoribbon on substrates with a different dielectric constant should allow one to 
tune accordingly the low-energy electronic states.

The paper is organized  as follows:  In Sec.~\ref{sec:model}, we define the model and introduce the projective QMC method. 
In Sec.~\ref{sec:static}, we focus on static properties of the system and address the behavior of equal-time  
spin and charge correlation functions upon increasing nonlocal Coulomb interactions. In Sec.~\ref{sec:dynamic},  
we analyze the corresponding evolution of single- and two-particle excitation spectra. 
Section~\ref{sec:summary} provides a summary and the conclusions.

\section{Model and the QMC method}
\label{sec:model}

We consider the following  Hamiltonian:
\begin{equation}
\hat{H}=-t\sum_{\langle \pmb{ij}\rangle,\sigma}
   \hat{c}^{\dag}_{{\pmb i}\sigma}\hat{c}^{}_{{\pmb j}\sigma} +
   \frac{1}{4} \sum_{\pmb{i}\pmb{j}}
  V_{\pmb{i}\pmb{j}}  ( \hat{n}_{\pmb{i}}   - 1 )  
  (  \hat{n}_{\pmb{j}}   - 1),
\label{eq:H}
\end{equation}
where $\hat{c}^{\dag}_{{\pmb i}\sigma}$ creates an electron with spin $\sigma$ at lattice site ${\pmb i}$ and 
$\hat{n}_{\pmb{i}}=\sum_{\sigma}\hat{c}^{\dag}_{{\pmb i}\sigma}\hat{c}^{}_{{\pmb i}\sigma}$ is the local particle 
number operator.  The first term $\hat{H}_t$ in Eq.~(\ref{eq:H}) describes the kinetic energy of the electrons 
with the hopping amplitude $t$ between nearest-neighbor sites $\langle \pmb{ij}\rangle$  on a finite-size 
half-filled  honeycomb lattice with open boundary conditions at the zigzag edges. 
The width $W$ of the graphene nanoribbon is defined by the number of zigzag chains with 2$L$ carbon atoms each 
and its length $L$ is measured in units of the lattice vector $\pmb{a}_{1}=(\sqrt{3}a,0)$ where $a$ is the carbon-carbon bond length. 
The second term $\hat{H}_V$ is a long-range density-density interaction with matrix elements,
\begin{equation}
  V_{\pmb{i}\pmb{j}} 
=   \left\{
    \begin{array}{ll}
      2 U\,, & \text{if  } | \pmb{i} - \pmb{j} | = 0 \\
      \frac{\alpha U a}{ | \pmb{i} - \pmb{j} | }  \,,  &  \text{if  } | \pmb{i}
      - \pmb{j} | >0.
    \end{array}
  \right.
\end{equation}
Here, $| \pmb{i} - \pmb{j} |$ is the minimal distance between two lattice sites $\pmb{i}$ and $\pmb{j}$  while  
$\alpha$ determines the relative strength of nonlocal interactions with respect to the on-site repulsion $U$.
In particular, for $\alpha=0$, one recovers the  Hubbard interaction:
\begin{equation}
  \hat{H}_U = \frac{U}{2} \sum_{\pmb{i}}\left(\hat{n}_{\pmb{i}} -1 \right)^2.
  \label{eq:Hub}
\end{equation}

To address the ground-state properties of the Hamiltonian (\ref{eq:H}), we use a projective (zero temperature) QMC algorithm 
based on the imaginary-time evolution of a trial wave function $| \Psi_\text{T}\rangle$ to the ground 
state $|\Psi_0 \rangle$,\cite{Assaad08_rev}                                   
\begin{equation}
  \frac{ \langle  \Psi_0 | \hat{O} |  \Psi_0 \rangle  }{ \langle  \Psi_0  |  \Psi_0 \rangle  }  =  
  \lim_{\Theta \rightarrow \infty} 
  \frac{ \langle  \Psi_\text{T} | e^{-\Theta \hat{H} /2 } \hat{O} e^{-\Theta \hat{H} /2  } |  \Psi_\text{T}\rangle  }
  { \langle  \Psi_\text{T}  | e^{-\Theta \hat{H} }  |  \Psi_\text{T} \rangle  }.
\end{equation}
It relies on a Trotter-Suzuki decomposition of the imaginary-time propagation, 
\begin{equation}
e^{-\Theta \hat{H} }= \prod_{i=1}^{L_\tau} e^{-\Delta\tau \hat{H}_t } e^{-\Delta\tau \hat{H}_V } + \mathcal{O}(\Delta\tau^2),
\end{equation}
into $L_{\tau}$ steps of size $\Delta\tau=\Theta/L_{\tau}$. Numerical results presented in Secs.~\ref{sec:static} and   
\ref{sec:dynamic} were obtained with the projection parameter $\Theta=60/t$, chosen sufficiently large to guarantee 
the convergence to the ground state $|\Psi_0\rangle$, and with an imaginary time discretization of $\Delta\tau=0.1/t$.

The simulations at finite $\alpha$ were performed by means of a recently implemented  generalization of the QMC 
method which allows one to handle long-range density-density interactions.~\cite{Hohenadler14} 
The approach we use, was introduced  and used in the context of  hybrid QMC methods.~\cite{Brower12,Ulybyshev13}  
It is based on the observation that the quartic nonlocal interaction term can be recast into a quadratic one 
by coupling the local electron density operator  to an auxiliary continuous scalar field, provided the interaction 
matrix $V_{\pmb{i}\pmb{j}}$ is positive definite. 
This renders it possible to integrate  out fermionic degrees of freedom at the expense of a determinant yielding 
an action solely dependent on the scalar field. The integration over the field configurations is carried out 
stochastically using the Monte Carlo method with a sequential  Metropolis updating scheme. 
Finally, a standard discrete Hubbard-Stratonovich transformation was employed in the limit of
a purely local Hubbard interaction.~\cite{Assaad08_rev} In order to ensure the explicit conservation 
of the SU(2) spin-rotation symmetry at half-filling, we have opted for an Ising spin auxiliary field coupling to the 
local charge operator at every space-time point, 
\begin{equation}
e^{-\Delta \tau  U \left( \hat{n}_{\pmb i} -  1   \right)^2   }  =   
\frac{1}{2}\sum_{s_{{\pmb i}\tau}=\pm 1} e ^{i \lambda s_{{\pmb i}\tau} \left( \hat{n}_{\pmb i} -  1   \right)},
\end{equation}
with coupling constant $\lambda$  given by  $\cos(\lambda) =   e^{-\Delta \tau  U/2}$.

\section{Static spin and charge correlations}
\label{sec:static}

Nonlocal Coulomb interactions can make charge fluctuations as large as the spin 
fluctuations.~\cite{Callaway89,Ohta94,Aichhorn04,Davoudi07,vLoon14,Huang14,Terletska17}  
This contrasts with systems with the repulsive on-site interaction only where the spin fluctuations dominate. 
In the extreme case, increasing the range of Coulomb interactions can result in a phase transition from a uniform state 
favoring singly occupied sites and hence, local moment formation, to the charge-ordered state with alternating 
doubly occupied and empty sites.

On the one hand, starting from the 1D and strong-coupling limits, the critical nearest-neighbor Coulomb 
repulsion driving the spin-density wave to a charge-density-wave phase transition along the 1D edge $V_c=U/2$  
is larger than $V_c=U/3$ in the bulk graphene with the coordination number 3. 
Thus, this oversimplified view implies robustness of the edge magnetism.  
On the other hand, Coulomb interaction couples edge and bulk states and the bulk corrections to the edge physics 
become more important with increasing ribbon width $W$.~\cite{Koop15}
This coupling makes the issue of the stability of magnetic edge correlations less obvious and might 
result in a complex many-body state with intertwined spin and charge fluctuations both contributing to 
the spectral properties of the nanoribbons.

Given this background, we perform QMC studies with the purpose of elucidating to what extent enhanced charge fluctuations 
affect the edge magnetism upon increasing the strength of nonlocal interactions controlled by  $\alpha$. 
The QMC simulations have been done at a fixed value of the on-site repulsion $U/t=2$, well below the transition from 
the semimetal to the bulk AF insulator at $U_c/t\simeq 3.8$ in the honeycomb Hubbard model,~\cite{Sorella12,Assaad13} 
and up to $\alpha_{\rm{max}}=1.23$ where charge and spin correlations are nearly degenerate in the classical limit.
Notably,  as estimated from previous QMC simulations, already $\alpha=1$ increases substantially the critical interaction 
$5<U_c/t<5.5$  for the aforementioned  magnetic transition signifying the importance of nonlocal 
interactions.~\cite{Hohenadler14}

\begin{figure}[t!]
\begin{center}
\includegraphics[width=0.48\textwidth]{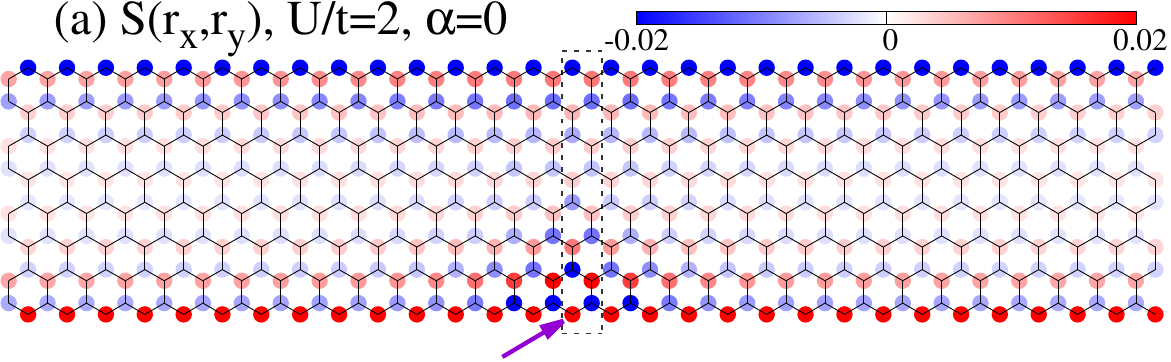}\\
\includegraphics[width=0.48\textwidth]{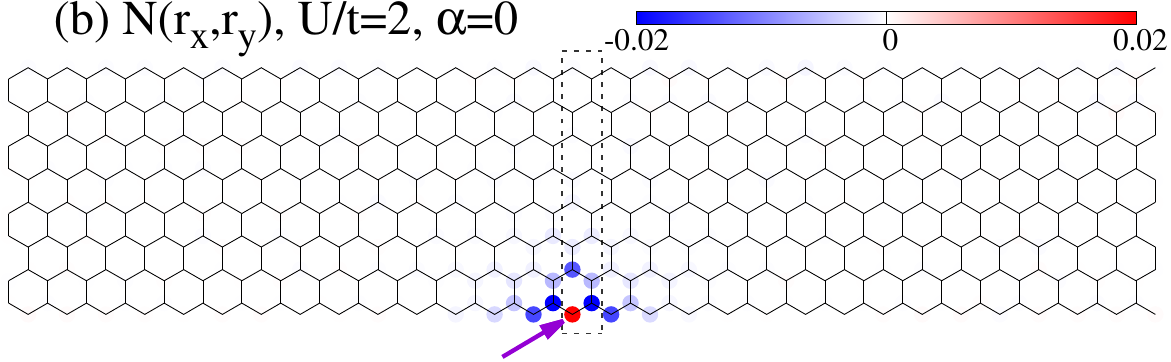}\\
\includegraphics[width=0.48\textwidth]{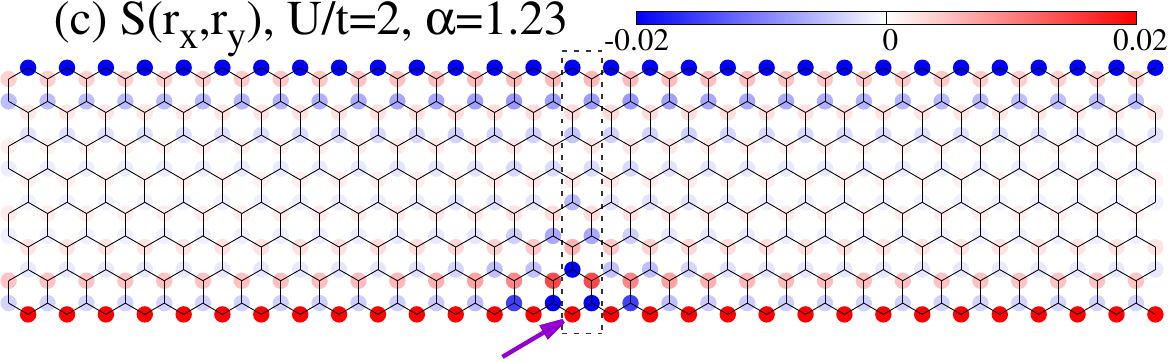}\\
\includegraphics[width=0.48\textwidth]{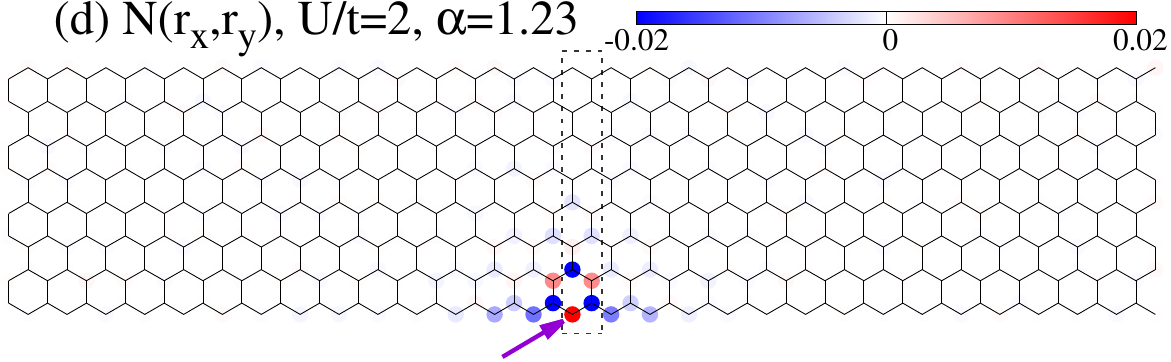}
\end{center}
\caption
{ 
Real-space spin $S(\pmb{r})$ and charge $N(\pmb{r})$ correlations relative to the central site of the edge 
indicated by the arrow for the Hubbard model ($\alpha=0$) (a,b)  and a $1/r$ Coulomb potential with $\alpha=1.23$ (c,d)  
from QMC simulations of a rectangular graphene nanoribbon, width $W=8$ and length $L=30$, corresponding to 
$N=2\times L\times W=480$ lattice sites. Dashed lines indicate the unit cell.  In both cases $U/t=2$.
}
\label{ribbon}
\end{figure}

First, we study properties of the ground state of the system. To this end, we calculate the static real-space
spin-spin correlation function,  
\begin{equation}
   S(\pmb{r})   = \frac{4}{3}\langle \pmb{\hat{S}}_{\pmb{r}} \cdot
  \pmb{\hat{S}}_{\pmb{0}}\rangle,
\end{equation}
and charge-charge correlation function,
\begin{equation}
   N(\pmb{r})   = \langle \hat{n}_{\pmb{r}}  \hat{n}_{\pmb{0}}\rangle -  \langle \hat{n}_{\pmb{r}}  \rangle  
  \langle \hat{n}_{\pmb{0}}\rangle.  
\end{equation}
The evolution of the real-space correlation pattern which stems from the QMC simulations of our largest nanoribbon with 
width $W=8$ and length $L=30$, measured relative to the central site of the edge (indicated by the arrow), 
is illustrated in Fig.~\ref{ribbon}. 
For a purely local interaction ($\alpha=0$)  one finds a clear evidence of quasi-long range FM spin 
correlations along the edges, see Fig.~\ref{ribbon}(a). The spin-spin correlations are antialigned to each other on the 
opposite ribbon edges and quickly decay into the inner sites.  The corresponding charge-charge correlations shown in 
Fig.~\ref{ribbon}(b) are slightly suppressed around the origin with respect to the noninteracting
system. Such behavior is expected for a fluid of charge carriers with contact interactions where charged particles 
avoid each other at short distances. 
In the case of a long-range interaction with $\alpha =1.23$, numerical data shown in Fig.~\ref{ribbon}(c,d) indicate  
the competition between short-range charge correlations,  driven by the nonlocal interactions, and AF spin correlations 
promoted by the on-site repulsion. Consequently, the spin correlations fall off into the bulk sites more quickly as 
compared to the Hubbard model, see Fig.~\ref{ribbon}(c). Nevertheless, the tendency towards the extended spin polarization
along the edges remains dominant over the sublattice charge fluctuations observed in $N(\pmb{r})$, see  
Fig.~\ref{ribbon}(d). 

As summarized in Table~\ref{table_d} for ribbons of width $W=6$ and 8, 
the enhancement of charge fluctuations upon increasing $\alpha$  is also signaled by a systematic growth of the average 
double occupancy in the system,
\begin{equation}
d=\frac{1}{N}\sum_{\pmb{i}}\langle \hat{n}^{}_{{\pmb i}\uparrow}\hat{n}^{}_{{\pmb i}\downarrow}\rangle.
\end{equation}
The increase of $d$ can be equally well  understood as resulting from the tendency of nonlocal Coulomb 
interactions to renormalize  downward the effective on-site interaction.~\cite{Schuler13}
Note that enhanced error bars at the smallest nonzero  $\alpha=0.25$  follow from particularly low acceptance 
rates within the adopted sequential Metropolis updating scheme of the time slices. 
However, the efficiency of the implemented sampling of the scalar field improves at larger $\alpha$ which is reflected 
in better data quality.

\begin{table}[b!]
\caption {Average double occupancy $d$ per lattice site upon increasing the strength of nonlocal interactions $\alpha$ 
for ribbons of width $W=6$ and 8.
}
\begin{ruledtabular}
\begin{tabular}{dll}
\multicolumn{1}{c}{}     &
\multicolumn{2}{c}{$d$}  \\
\multicolumn{1}{c}{$\alpha$} &
\multicolumn{1}{c}{$W=6$}    &
\multicolumn{1}{c}{$W=8$}  \\
\hline
0         & 0.19388(2)  &  0.19533(2) \\
0.25      & 0.2004(1)   &  0.2016(2)  \\
0.5       & 0.20732(3)  &  0.20862(7) \\
0.75      & 0.21467(3)  &  0.21595(3) \\
1         & 0.22250(2)  &  0.22373(2) \\
1.23      & 0.23014(2)  &  0.23133(1) \\
\end{tabular}
\end{ruledtabular}
\label{table_d}
\end{table}

\begin{figure}[t!]
\begin{center}
\includegraphics[width=0.48\textwidth]{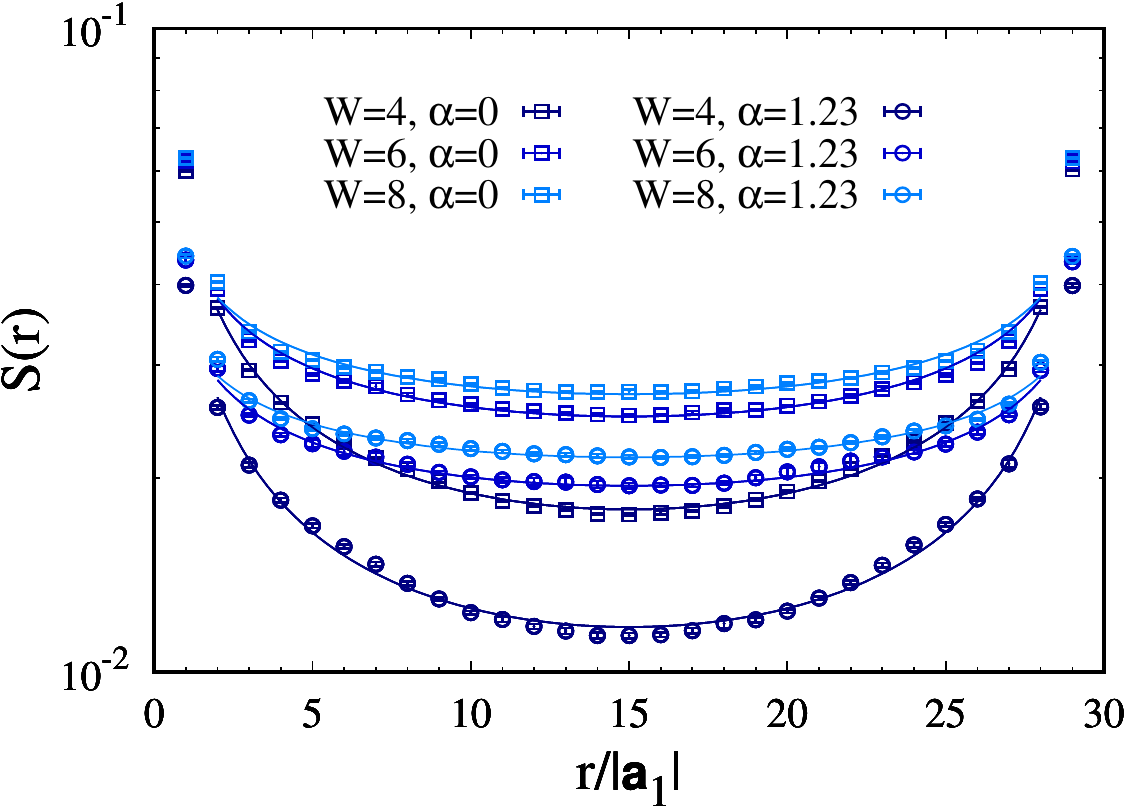}
\end{center}
\caption
{Spin-spin correlation function $S(r)$  along the zigzag edge for nanoribbons with different width $W$ 
in the Hubbard model ($\alpha=0$, squares)  and for  a $1/r$ Coulomb potential ($\alpha=1.23$, circles).
}
\label{edge}
\end{figure}

In order to address the edge magnetism  in the presence of long-range Coulomb interactions more quantitatively, 
we plot in Fig.~\ref{edge} the spin-spin correlation function $S(r)$ as a function of the distance $r$ along the edge. 
Following up Ref.~\onlinecite{Feldner11}, we  consider a function,    
\begin{equation}
 S_{\text{fit}}(r)\propto r^{-\eta} e^{-r/\xi} + (L-r)^{-\eta} e^{-(L-r)/\xi},
 \label{S_fit}
\end{equation}
combining both exponential and power-law behaviors.
An important conclusion of the QMC studies in Ref.~\onlinecite{Feldner11} reached on systems up to $L=60$ was that the 
correlation length $\xi$  increases with the width $W$ of ribbons such that already the data for the $W=6$ ribbons 
can be equally well fitted assuming an infinite spin correlation length. 
By fitting $S_{\text{fit}}(r)$ to our QMC data obtained on shorter systems with $L=30$, we find exceedingly 
large magnetic correlation length  even for the narrowest ribbon with $W=4$. However, as summarized in Table~\ref{table}, 
the actual change of exponent $\eta$ with increasing strength of the nonlocal part of Coulomb interactions $\alpha$ 
is the width-dependent.

In the case of the narrow $W=4$ ribbons, finite $\alpha$ enhances the interedge AF coupling and brings the system 
closer to the regime of even-leg spin ladders as signals by a larger with respect to $\alpha=0$ exponent $\eta$ 
and thus a more rapid exponential-like decay of spin correlations.~\cite{White94,Dagotto96} 
The tendency to form rung singlets becomes less important for larger interedge distances. 
Already for the $W=6$ ribbons, the dominant effect is the reduction of short-range spin correlations by 
competing charge-density fluctuations while leaving the long-wavelength physics nearly unaltered. 
This is seen as an effective reduction of the exponent $\eta$ with $\alpha$, cf. Table~\ref{table}. 
Same behavior is also resolved for the $W=8$ ribbons.

\begin{figure}[t!]
\begin{center}
\includegraphics[width=0.48\textwidth]{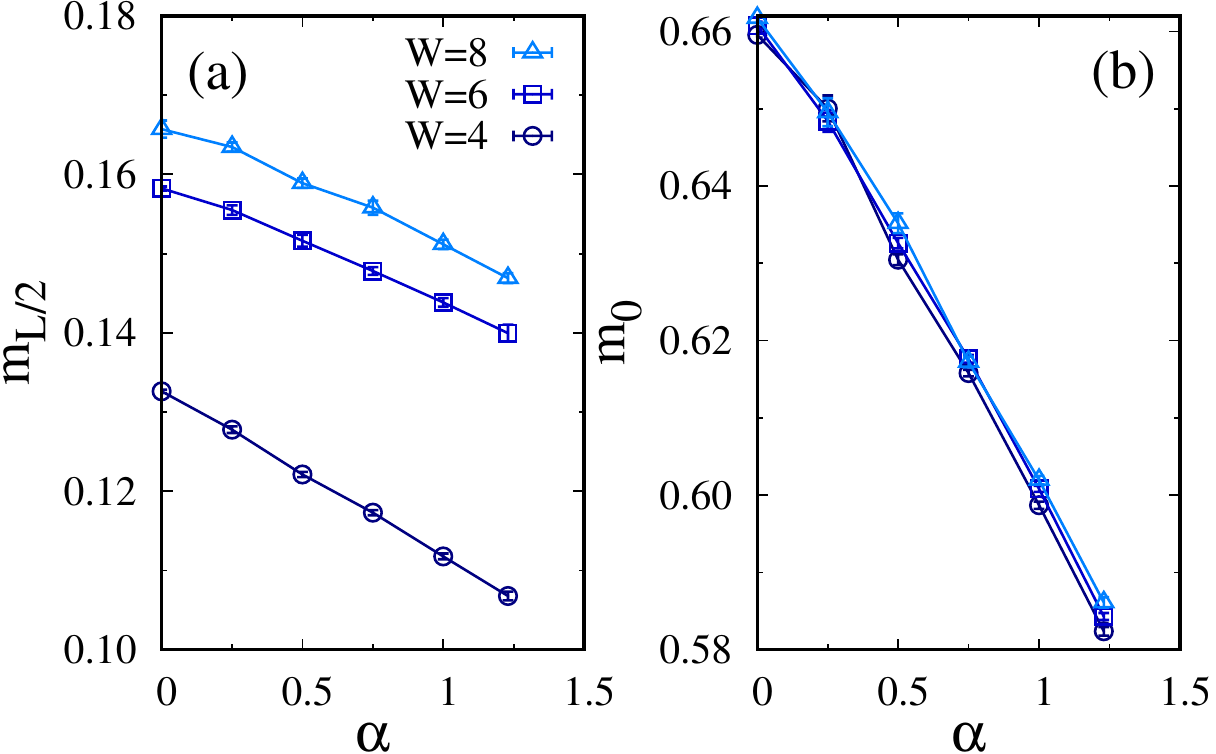}
\end{center}
\caption
{(a) Magnetic moment at the largest distance  $m_{L/2}$  and (b) local moment $m_0$ at the ribbon edge as 
a function of the strength of  nonlocal interactions $\alpha$.
}
\label{edge2}
\end{figure}

\begin{table}[b!]
\caption{ 
Exponent $\eta$ as obtained by fitting the function $S_{\text{fit}}(r)$ to the QMC data for ribbons 
with different width $W$.
}
\begin{ruledtabular}
\begin{tabular}{lccc}
                &     $W=4$   &    $W=6$    &    $W=8$     \cr
\colrule
 $\alpha=0$     &     0.61(1) &     0.41(1) &    0.35(1)   \cr
 $\alpha=1.23$  &     0.67(1) &     0.37(1) &    0.30(1)   \cr
\end{tabular}
\end{ruledtabular}
\label{table}
\end{table}

The special character of the $W=4$ ribbon with the relatively strongest reduction of spin correlations 
upon increasing $\alpha$ is further confirmed by comparing in Fig.~\ref{edge2}(a) the evolution of the magnetic 
moment at the largest distance along the edge,
\begin{equation}
m_{L/2}=\sqrt{\frac{4}{3}\langle  \pmb{\hat{S}}_{L/2}\cdot\pmb{\hat{S}}_{0} \rangle}, 
\end{equation}
with that obtained for wider $W=6$ and 8 ribbons. In contrast, the reduction of the local moment at the edge,
\begin{equation}
m_0=\frac{4}{3}\langle \pmb{\hat{S}}_0^2\rangle = 1 - 2\langle \hat{n}_{0\uparrow}\hat{n}_{0\downarrow} \rangle,
\label{m_vs_d}
\end{equation}
is very much the same for all the ribbon geometries studied in this work, see Fig.~\ref{edge2}(b).

\begin{figure}[t!]
\begin{center}
\includegraphics[width=0.48\textwidth]{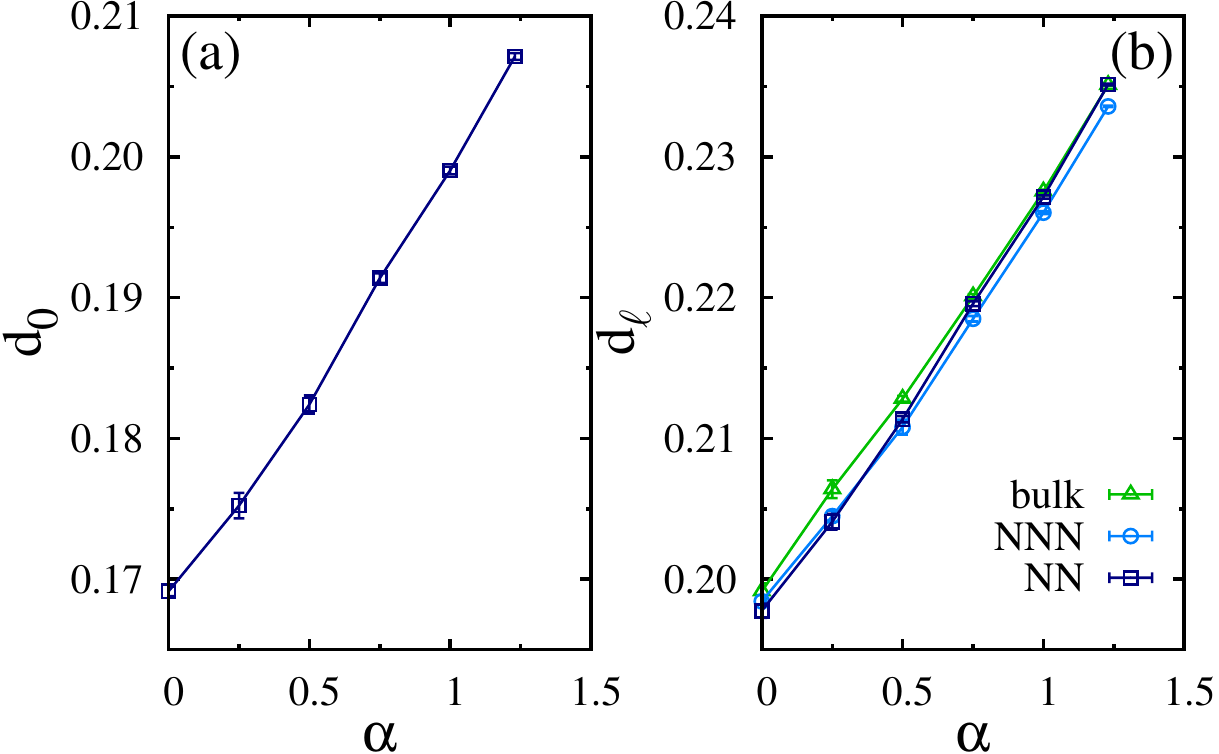}
\end{center}
\caption
{(a) Double occupancy $d_0$ at the ribbon edge and (b) double occupancy $d_{\ell}$ at its nearest-neighbor (NN) and  
next-nearest-neighbor (NNN) sites in the unit cell as well as in the bulk upon increasing the strength of nonlocal interactions $\alpha$.
The ribbon width  $W=8$. Note the difference in the scales of vertical axis between both panels.  
}
\label{d_spatial}
\end{figure}

Equation (\ref{m_vs_d}) implies that  a decrease in the local moment is equivalent to the enhancement of double 
occupancy and reflects growing charge fluctuations in the system.  Thus, we study next the evolution of spatial dependence 
of double occupancy $d_{\ell}$ upon increasing $\alpha$. To this end, we plot in Fig.~\ref{d_spatial} double occupancy 
for the widest $W=8$  ribbon at the edge as well as at its nearest-neighbor (NN) and  next-nearest-neighbor (NNN) sites 
in the unit cell. For comparison, we also show double occupancy in the bulk. 
On the one hand, due to the tendency to local moment formation, the edge  clearly stands out from the rest through the strongest 
reduction of double occupancy, see Fig.~\ref{d_spatial}(a).  On the other hand, double occupancy of both the NN and NNN sites 
is essentially indistinguishable from the bulk and becomes at our largest $\alpha=1.23$ only slightly reduced compared to  
its noninteracting value 0.25,  see Fig.~\ref{d_spatial}(b).

\section{Dynamical spectral functions} 
\label{sec:dynamic}

Given the tendency of nonlocal Coulomb interactions to renormalize downwards the effective on-site interaction,~\cite{Schuler13} 
the reduction of spin correlations along the edge by nonlocal interactions revealed in Sec.~\ref{sec:static} 
can be mimicked within a simplified model with a purely local interaction. However, an important question 
is how accurately the actual physical situation in the system can be accounted for by such an effective model?

To address this issue let us now discuss the single- and two-particle excitation spectra of the system. 
Using the translational invariance of the system along the edge direction, it is convenient to introduce 
a momentum-space representation of the fermion operator,
\begin{equation} 
\hat{c}_{\ell q\sigma}=\frac{1}{\sqrt{L}} \sum_{j=1}^{L} e^{i q |\pmb{a}_1|  j} \hat{c}_{i(j,\ell)\sigma}, 
\end{equation}
where the lattice site $i(j,\ell)$  is given by the unit cell index  $j$ and the intracell index $\ell$.  
The knowledge of the momentum-resolved Green's function,  
\begin{equation}
G_\ell(q,\tau)=
\frac{1}{2}\sum_{\sigma}\langle \hat{c}^{}_{\ell q\sigma}(\tau) \hat{c}^{\dag}_{\ell q\sigma}(0)\rangle, 
\end{equation}
allows one to obtain the single-particle spectral function on the real frequency axis $\omega$ by the inversion of,
\begin{equation}
   G_\ell(q,\tau)=  \frac{1}{\pi} \int{\rm d}\omega  \; e^{-\tau\omega} A_{\ell}(q,\omega),
\end{equation}
by means of the stochastic analytic continuation method.~\cite{Beach04a} 
Similarly, imaginary-time-displaced  spin-spin and charge-charge  correlation functions
allow one to extract the dynamical spin $S_{\ell}(q, \omega)$   and charge $C_{\ell}(q, \omega)$ 
structure factors by analytic continuation of,
\begin{align}
        \frac{4}{3}\langle  \pmb{\hat{S}}_{\ell}(q,\tau) \cdot \pmb{\hat{S}}_{\ell}(-q,0) \rangle 
   & = \frac{1}{\pi}\int {\rm d} \omega  \;    e^{-\tau \omega } S_{\ell}(q, \omega), \\
                   \langle  \hat{N}_{\ell}(q,\tau) \cdot \hat{N}_{\ell}(-q,0) \rangle 
   & = \frac{1}{\pi}\int {\rm d} \omega  \;    e^{-\tau \omega } C_{\ell}(q, \omega), 
\end{align}
where, 
\begin{align} 
 \pmb{\hat{S}}_{\ell}(q)  & = 
 \frac{1}{\sqrt{L} } \sum_{j=1}^{L} e^{i q |\pmb{a}_1|  j} \pmb{\hat{S}}_{i(j,\ell)},  \\ 
       \hat{N}_{\ell}(q)  & = 
  \frac{1}{\sqrt{L} } \sum_{j=1}^{L}  e^{i q |\pmb{a}_1|  j}
\left( \hat{n}_{i(j,\ell)}   - n\right), 
\end{align}
and $n$ is the average filling level. Below we discuss dynamical spectral functions along the ribbon edge and 
within the center (bulk) obtained for our largest ribbon of width $W=8$.

\subsection{Single-particle spectral function}
\label{sec:single}

\begin{figure}[t!]
\begin{center}
\includegraphics[width=0.45\textwidth]{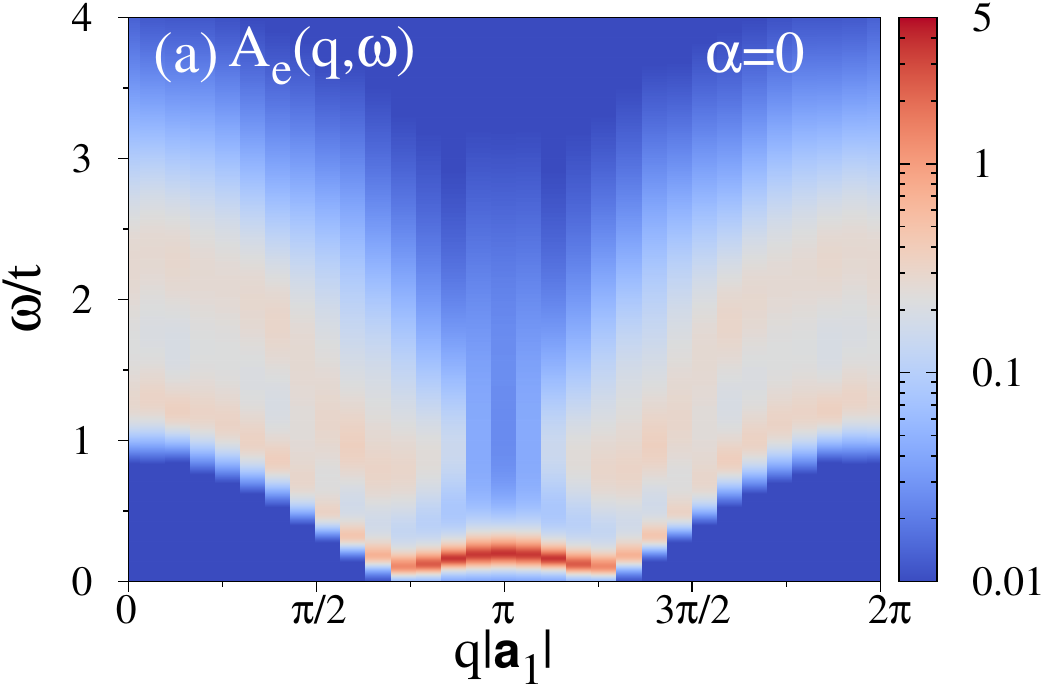}\\
\includegraphics[width=0.45\textwidth]{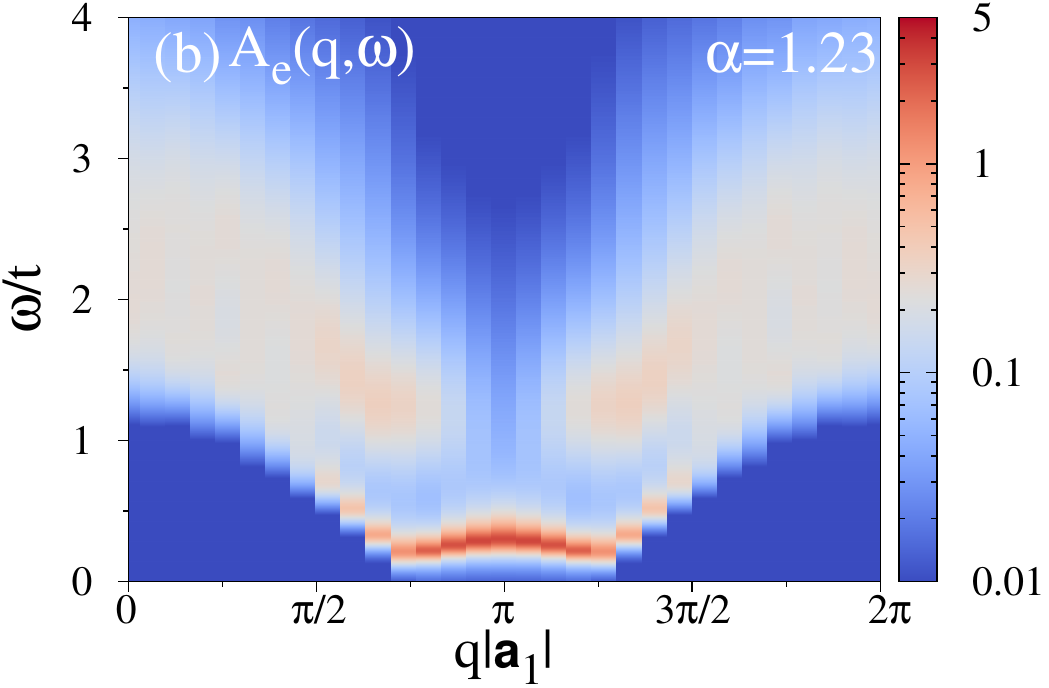}
\end{center}
\caption
{
Momentum resolved single-particle spectral function along the zigzag edge $A_e(q,\omega)$  for a ribbon 
of width $W=8$ with $\alpha=0$ (a)  and $\alpha=1.23$ (b). Color scheme for intensity plots from 
Ref.~\onlinecite{gnu}.
}
\label{Akw}
\end{figure}

The essential feature of the  tight-binding ($U=0$) band structure of a semi-infinite graphene sheet with a zigzag edge  
is a pair of flat zero-energy  bands in the range   $\tfrac{2\pi}{3} \le q|{\pmb a}_1| \le \tfrac{4\pi}{3}$ 
between the two Dirac points. Electron interactions modify the single-particle  spectrum and lead to distinct 
spectral features related to the onset of edge magnetism. 
We illustrate this effect in Fig.~\ref{Akw}. It  compares  a momentum resolved single-particle spectral function
along the zigzag edge  $A_e(q,\omega)$   in the limit of local ($\alpha=0$) Hubbard interaction [Fig.~\ref{Akw}(a)]  
with that obtained with nonlocal interactions $\alpha=1.23$ [Fig.~\ref{Akw}(b)].
Note that the considered value $U/t=2$ is chosen below the critical value for the Mott transition on the honeycomb lattice
$U_c/t\simeq 3.8$.~\cite{Sorella12,Assaad13} Hence, on our limited-size lattice, the gap around $q|{\pmb a}_1|=\tfrac{3\pi}{4}$ 
and  $q|{\pmb a}_1|=\tfrac{5\pi}{4}$ wavevectors is a finite size effect. 
Moreover, the observed in Ref.~\onlinecite{Hohenadler14} increase of the critical value $U_c$ upon going from a Hubbard 
to a long-range interaction, ensures the gapless nature of edge modes such that the system remains a semimetal.

\begin{figure}[t!]
\begin{center}
\includegraphics[width=0.45\textwidth]{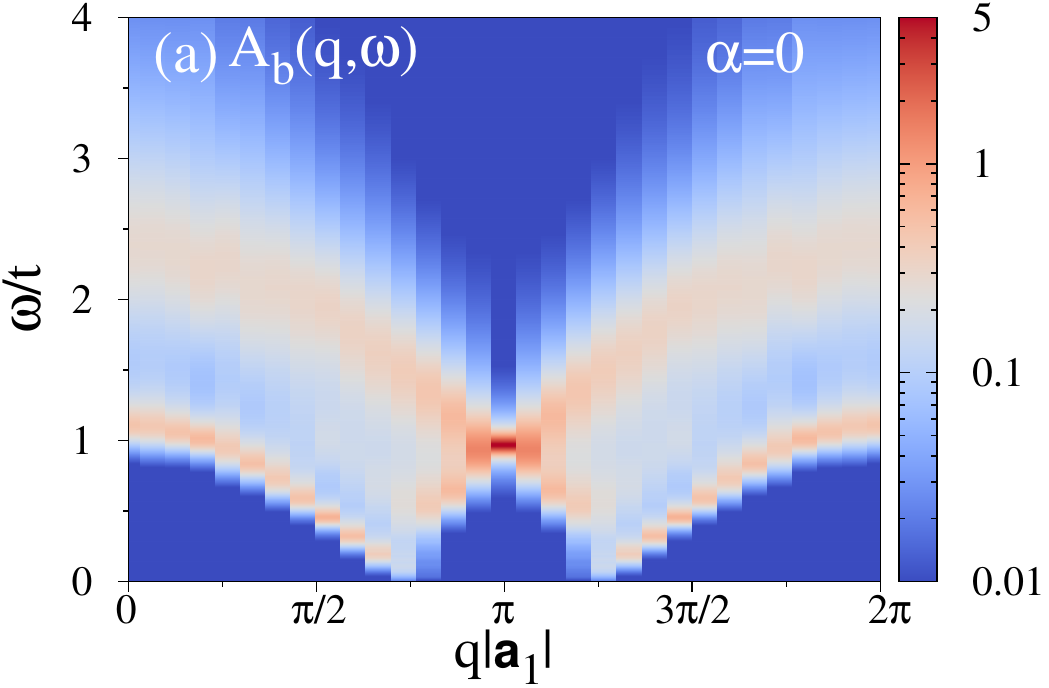}\\
\includegraphics[width=0.45\textwidth]{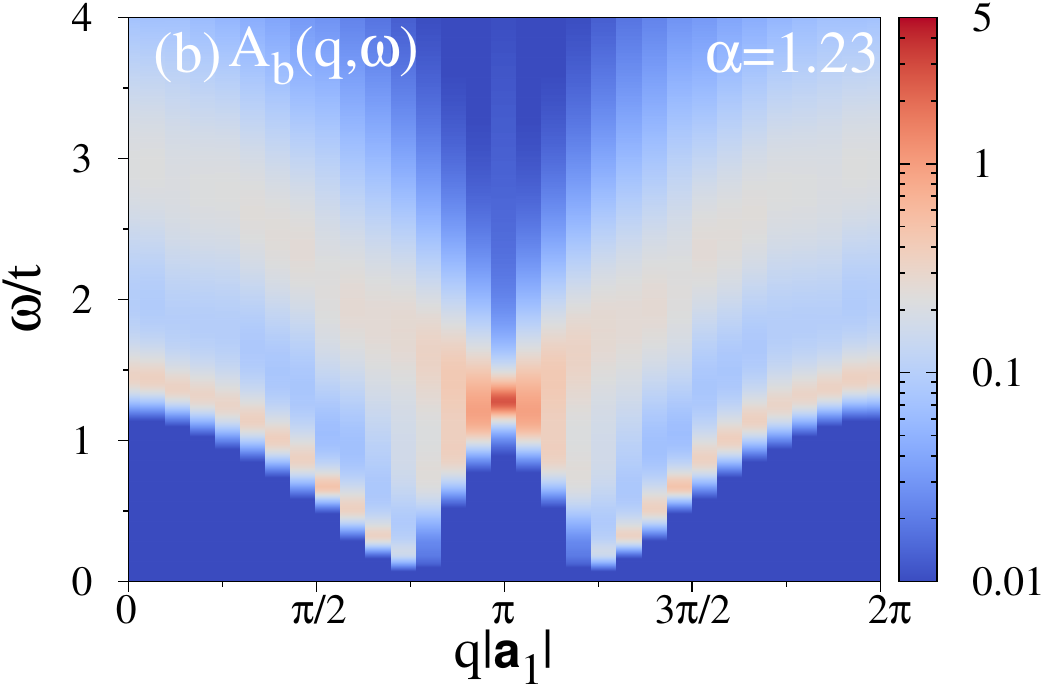}
\end{center}
\caption
{
Same as in Fig.~\ref{Akw} but for the bulk sites. 
}
\label{Akw_b}
\end{figure}

\begin{figure}[b!]
\begin{center}
\includegraphics[width=0.48\textwidth]{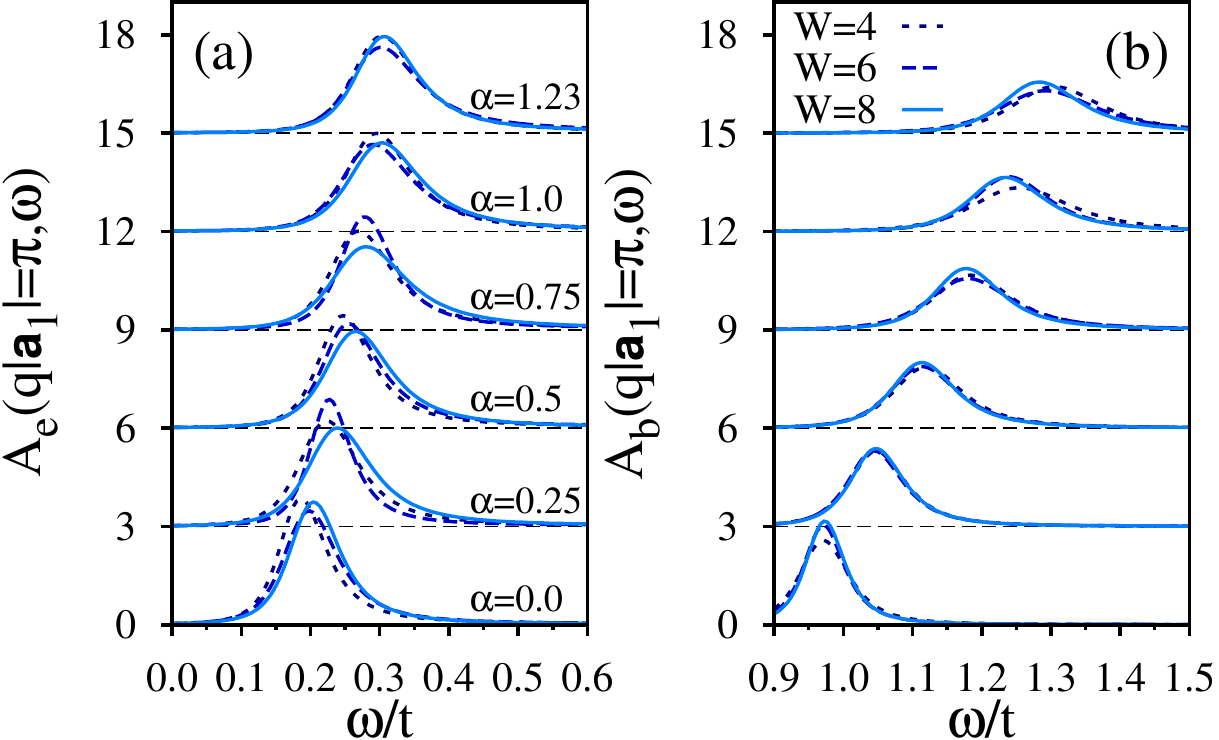}
\end{center}
\caption
{
Evolution of the single-particle spectrum at the edge $A_e(q|{\pmb a}_1|=\pi,\omega)$ (a) and 
in the bulk $A_b(q|{\pmb a}_1|=\pi,\omega)$ (b) upon increasing the strength of nonlocal interactions $\alpha$ 
for ribbons of different  widths $W$. The data in panel (b) were scaled down by a factor 0.6. 
}
\label{gap}
\end{figure}

As apparent, in the presence of the Hubbard  interaction [Fig.~\ref{Akw}(a)], the initially flat edge-localized states  
acquire a finite dispersion reaching its maximum at the $q|{\pmb a}_1|=\pi$ momentum.  The position of the single-particle 
peak at the $q|{\pmb a}_1|=\pi$ wavevector  is proportional to both the strength of the Hubbard interaction 
and the magnetic moment at the edge sites,~\cite{Feldner11} $\Delta_{\rm {sp}}(q|{\pmb a}_1|=\pi)\propto Um_{L/2}$. 
As shown in Fig.~\ref{Akw}(b),  switching on nonlocal interactions renormalizes further the dispersion of edge states 
and gives rise to the enhancement of its bandwidth.  
While it is suggestive of the many-body renormalization of the electron velocity at low energies found in bulk graphene,~\cite{Elias11} 
we note that much larger system sizes than those studied in this work are required to unambiguously resolve 
the logarithmically divergent Fermi velocity for edge modes near the Dirac points.~\cite{Affleck17} 

In Fig.~\ref{Akw_b} we display the corresponding single-particle spectrum in the bulk $A_b(q,\omega)$. 
A complete suppression of the low-energy spectral weight for wavevectors  
$\tfrac{3\pi}{4}<q|{\pmb a}_1|<\tfrac{5\pi}{4}$ can be traced back to the localized nature of the edge states 
which quickly decay into the inner sites. 
Furthermore, upon going from a Hubbard [Fig.~\ref{Akw_b}(a)] to a long-range interaction [Fig.~\ref{Akw_b}(b)], 
one observes a substantial widening of the single-particle spectrum. 

To get more insight into this effect and, in particular, to ascertain whether the observed renormalization of 
the single-particle spectral function at the ribbon edge and in the bulk can be traced to the same origin, 
we focus on the $q|{\pmb a}_1|=\pi$ momentum.  This wavevector is of major relevance to magnetic properties of the system:   
it corresponds to the most localized states along the edge and thus---according to 
the Stoner criterion---plays a dominant role in the spontaneous appearance of FM spin correlations 
along the edges. 

A detailed evolution of $A_e(q|{\pmb a}_1|=\pi,\omega)$ and $A_b(q|{\pmb a}_1|=\pi,\omega)$ 
upon increasing $\alpha$ for a ribbon of width $W=8$   is summarized in Fig.~\ref{gap}. 
We also plot the spectral data from the QMC simulations on narrower $2\times 30\times 4$ and $2\times 30\times 6$ 
ribbons.
First of all, the observed transfer of spectral weight towards higher frequencies turns out to be robust 
against changes in the nanoribbon width excluding the possibility that it simply stems from a finite size effect  
enhanced by long-range Coulomb interactions. Furthermore,  the striking difference between both panels is 
a steady shift in the peak position in $A_b(q|{\pmb a}_1|=\pi,\omega)$  which persists even at our largest $\alpha=1.23$. 
Thus, one may conclude that the shift originates from the long-range tail of Coulomb interactions;
the resultant renormalization of the electron velocity at the Dirac points produces a rigid shift of the 
electronic structure in the bulk towards higher energies, see Fig.~\ref{Akw_b}.

\begin{figure}[t!]
\begin{center}
\includegraphics[width=0.45\textwidth]{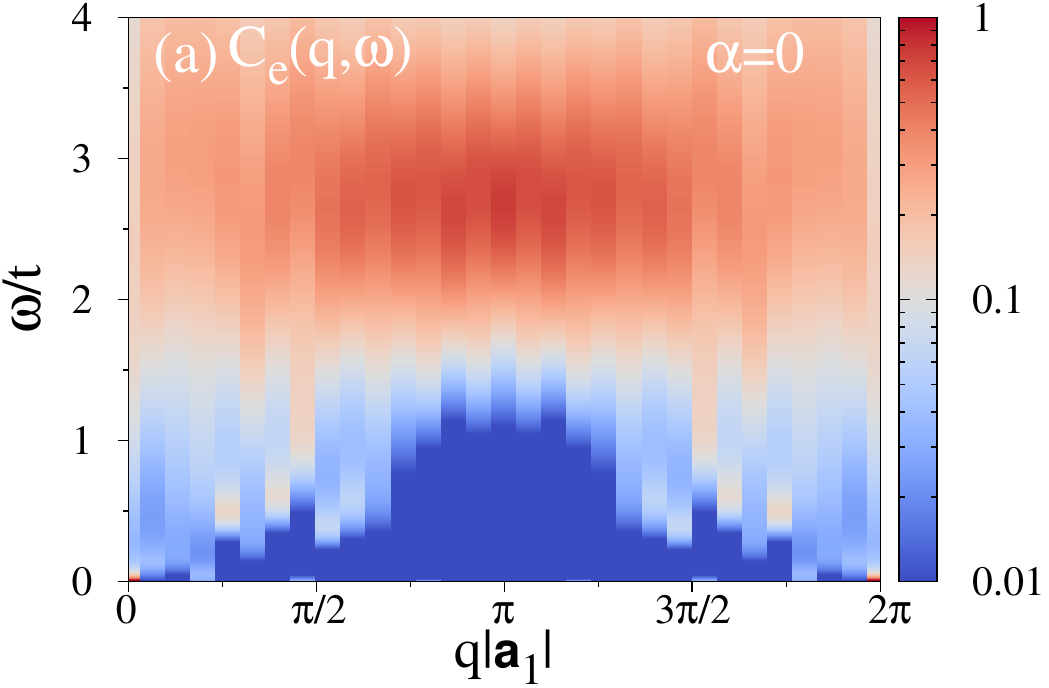}\\
\includegraphics[width=0.45\textwidth]{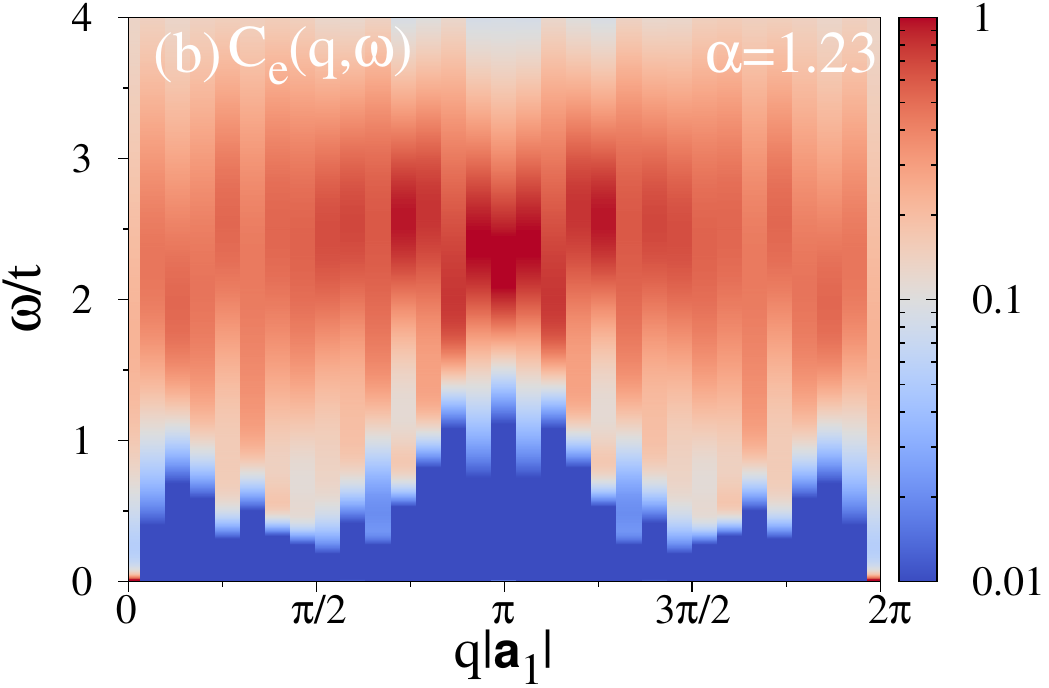}
\end{center}
\caption
{
Dynamical charge structure factor along the zigzag edge $C_e(q,\omega)$  for a ribbon of width $W=8$ with 
$\alpha=0$ (a)  and $\alpha=1.23$ (b).
}
\label{Ckw}
\end{figure}

\begin{figure}[t!]
\begin{center}
\includegraphics[width=0.45\textwidth]{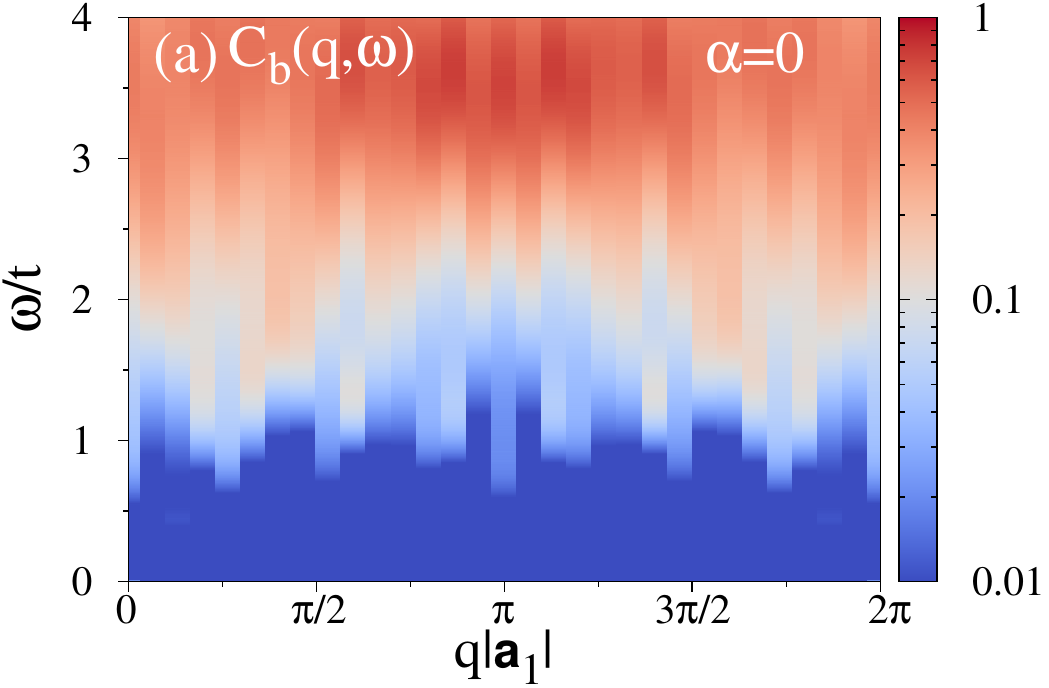}\\
\includegraphics[width=0.45\textwidth]{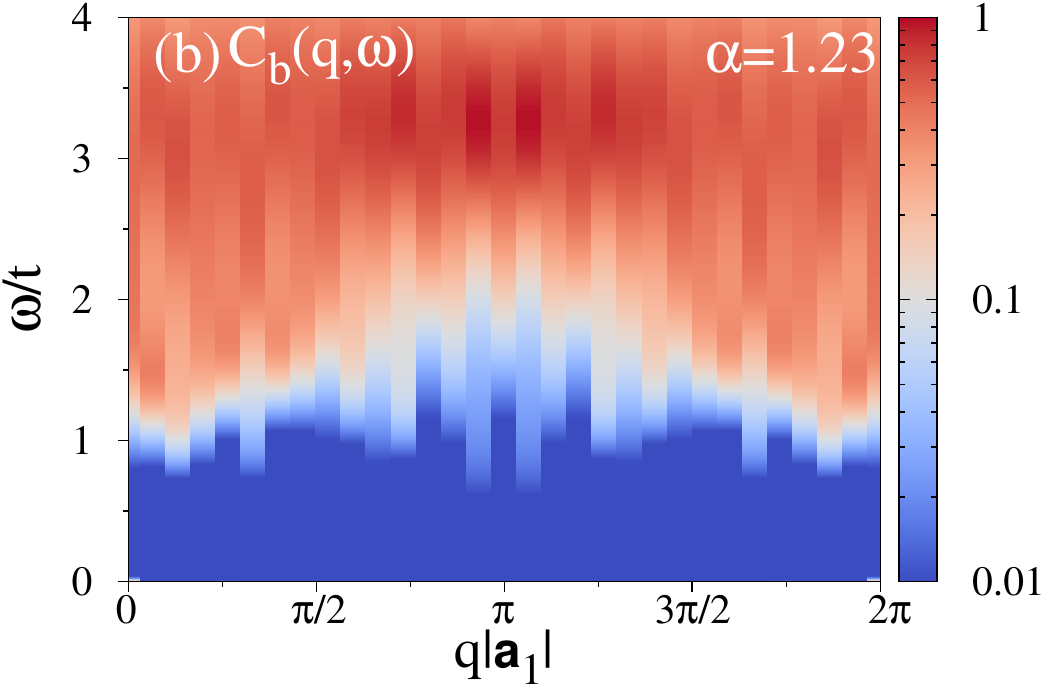}
\end{center}
\caption
{
Same as in Fig.~\ref{Ckw} but for the bulk sites.
}
\label{Ckw_b}
\end{figure}

In contrast, the corresponding transfer of spectral weight in the single-particle spectrum at the edge 
$A_e(q|{\pmb a}_1|=\pi,\omega)$ seems to saturate with increasing $\alpha$. This indicates that it stems from 
short-range interactions. 
Such an interpretation is further supported by the results of Ref.~\onlinecite{Affleck17} reporting a significant change 
in the position of the low-energy peak $\Delta_{\rm {sp}}(q|{\pmb a}_1|=\pi)$ in the presence of Coulomb interactions 
with a short-range real-space cutoff.   
However, recalling that in the Hubbard model, $\Delta_{\rm {sp}}(q|{\pmb a}_1|=\pi)\propto Um_{L/2}$, 
the simultaneous shift of $\Delta_{\rm {sp}}(q|{\pmb a}_1|=\pi)$ towards higher frequencies and the reduction 
of the magnetic moment $m_{L/2}$  cannot be consistently reproduced in the effective model with a renormalized on-site  
interaction thus pointing  out the intricate nature of the dynamics of edge states.

\subsection{Dynamical charge structure factor}
\label{sec:charge}

In order to resolve this puzzle, we turn now our attention to the charge dynamics. From the equal-time data in 
Fig~\ref{ribbon}(d), revealing  enhanced short-range charge correlations at finite $\alpha$, one might expect to resolve 
the concomitant redistribution of charge spectral weight in the dynamical charge structure factor 
along the zigzag edge $C_e(q,\omega)$.  The comparison of $C_e(q,\omega)$ obtained in the Hubbard model 
[Fig.~\ref{Ckw}(a)] with that calculated with long-range interactions [Fig.~\ref{Ckw}(b)] does 
indeed reveal the emergence of broad incoherent low-energy excitations. The observed softening near 
$q|{\pmb a}_1|=\tfrac{\pi}{2}$ and $q|{\pmb a}_1|=\tfrac{3\pi}{2}$ wavevectors stems from scattering between the Dirac 
points and reflects growing charge fluctuations in the system. 
Therefore,  it is natural to ascribe the shift of $\Delta_{\rm {sp}}(q|{\pmb a}_1|=\pi)$ in Fig.~\ref{gap}(a) to 
short-range charge-density-wave fluctuations. In this scenario, the quasiparticle scattering off charge fluctuations 
adds up to scattering off  spin fluctuations and effectively reinforces the edge magnetism induced single-particle 
gap at the $q|{\pmb a}_1|=\pi$ wavevector. 
Finally, we also note that the quickly vanishing low-energy charge weight with increasing momentum $q$ in Fig.~\ref{Ckw}(b) 
is consistent with the expected increase of the velocity of long-wavelength charge excitations in the presence 
of nonlocal interactions.

\begin{figure}[t!]
\begin{center}
\includegraphics[width=0.45\textwidth]{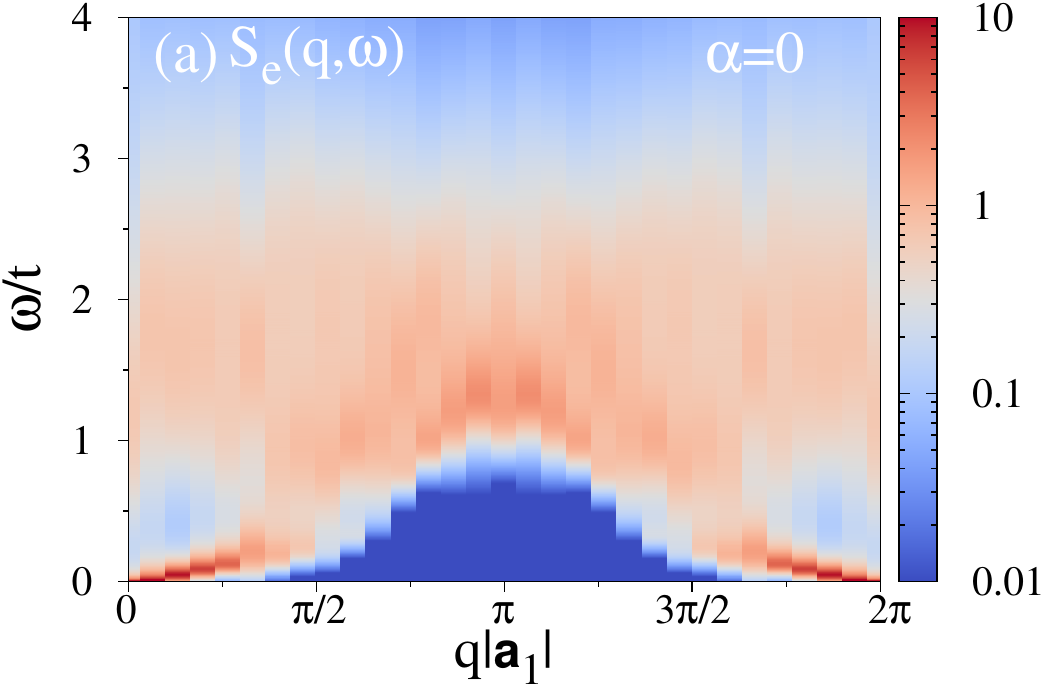}\\
\includegraphics[width=0.45\textwidth]{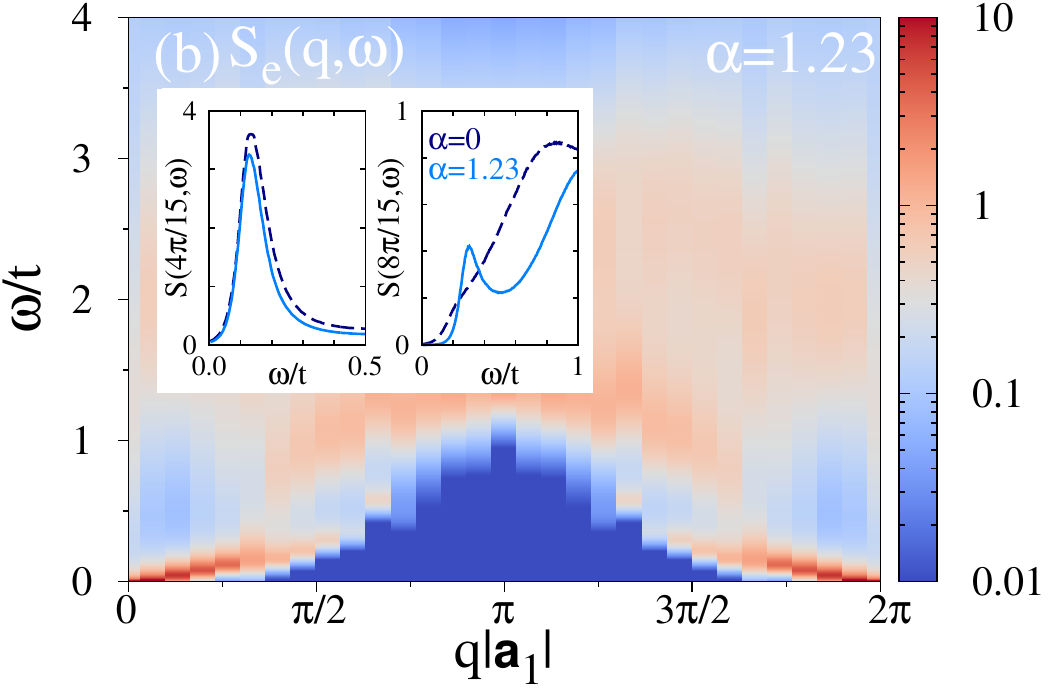}
\end{center}
\caption
{
Dynamical spin structure factor along the zigzag edge $S_e(q,\omega)$  for a ribbon of width $W=8$ with 
$\alpha=0$ (a) and $\alpha=1.23$ (b).  Inset shows the low-energy part of the spectrum at 
$q|{\pmb a}_1|=\frac{4\pi}{15}$ (left)  and $q|{\pmb a}_1|=\frac{8\pi}{15}$ (right) for $\alpha=0$ (dashed) 
and $\alpha=1.23$ (solid). While $S_e(q,\omega)$ remains essentially identical for small momenta, the spin-wave-like 
excitations continue to disperse up to larger wavevectors in the presence of nonlocal interactions.  
}
\label{Skw}
\end{figure}

\begin{figure}[t!]
\begin{center}
\includegraphics[width=0.45\textwidth]{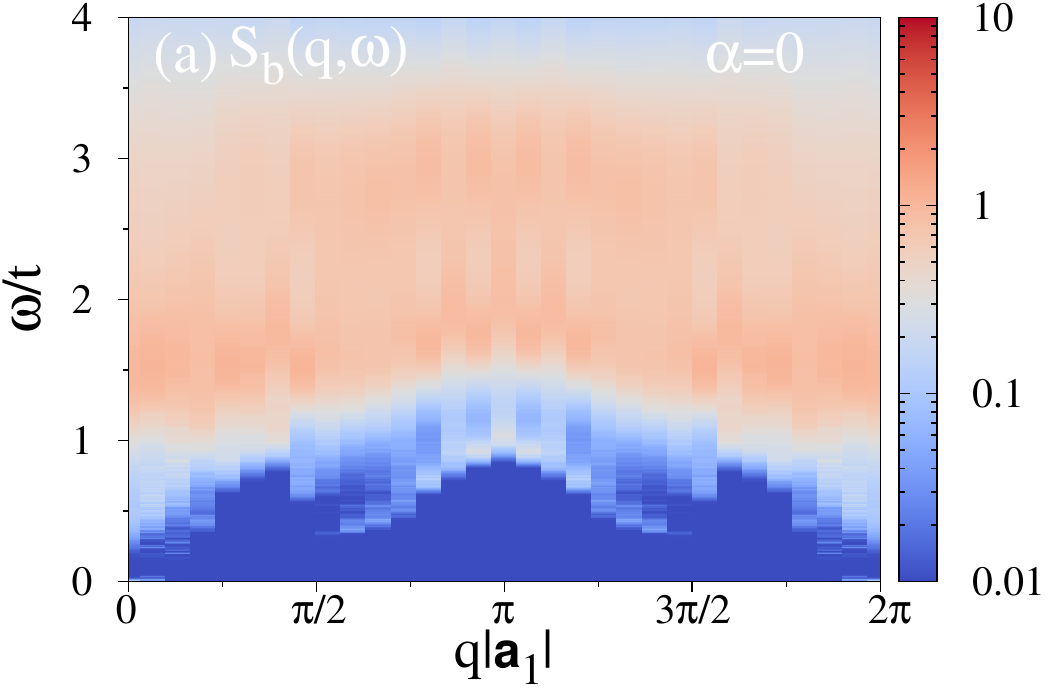}\\
\includegraphics[width=0.45\textwidth]{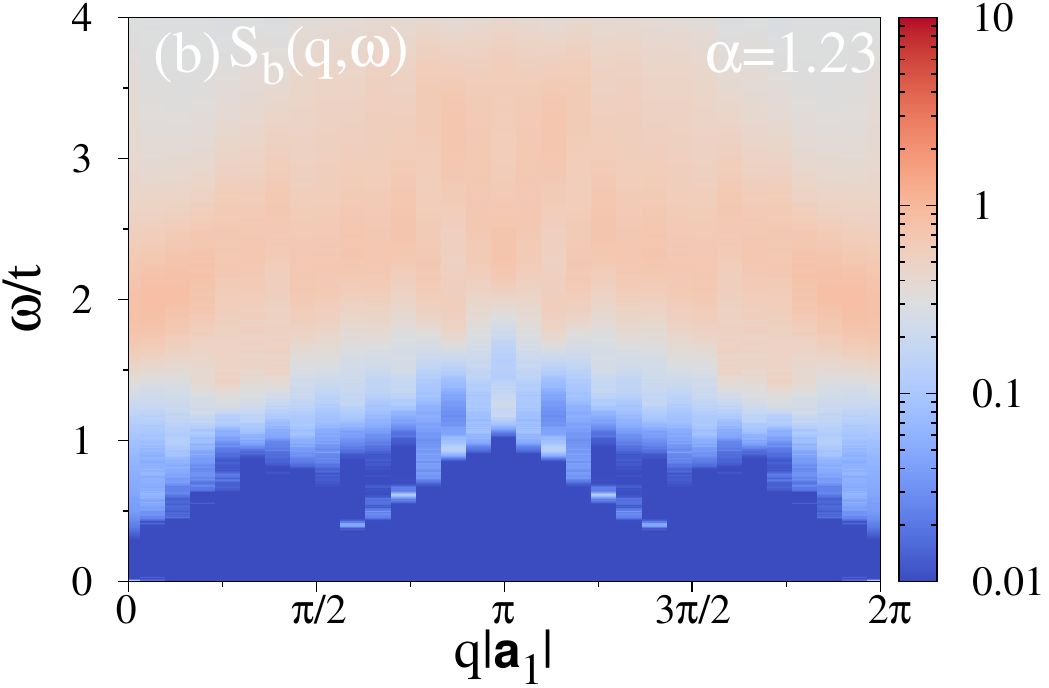}
\end{center}
\caption
{
Same as in Fig.~\ref{Skw} but for the bulk sites.
}
\label{Skw_b}
\end{figure}

Given the semimetallic nature of the bulk, it is expected to retrace soft features seen in $C_e(q,\omega)$ 
in the charge excitation spectrum in the bulk. However, the corresponding dynamical charge structure factor $C_b(q,\omega)$ 
in Fig.~\ref{Ckw_b} appears to be gapfull. The apparent contradiction is resolved by comparing the single-particle 
spectra along the edge $A_e(q,\omega)$ (Fig.~\ref{Akw})  and in the bulk $A_b(q,\omega)$ (Fig.~\ref{Akw_b}).  
The presence of a weakly dispersive  quasiparticle band along the edge directly affects 
the low-energy scattering between the Dirac points. The resulting large phase space for particle-hole  
excitations gives rise to  an extended particle-hole continuum in the dynamical charge structure factor along the 
edge  $C_e(q,\omega)$, see Fig.~\ref{Ckw}(b). 
In contrast, a pointlike Fermi surface with vanishing density of states in the bulk strongly restricts decay channels 
for low-energy quasiparticles due to phase space limitations. 
Consequently,  the intensity of the particle-hole continuum quickly decreases in the low-energy limit making it very 
difficult to resolve the corresponding charge excitations in $C_b(q,\omega)$, see  Fig.~\ref{Ckw_b}(b).

\subsection{Dynamical spin structure factor}
\label{sec:spin}

Finally, let us address the effect of nonlocal Coulomb interactions on the spin dynamics. 
Figure~\ref{Skw}(a) reports the dynamical spin structure factor along the edge $S_e(q,\omega)$ 
in the Hubbard model. As apparent, a prominent signature of quasi-long range FM spin correlations 
is the roughly linear low-energy spin-wave-like mode which merges into a particle-hole continuum at higher frequencies.
Considering the FM nature of spin correlations along the edge, it might---at first glance---be surprising
to see that the spin-wave dispersion is not quadratic in the long-wavelength limit. 
However, the linear dispersion is a consequence of AF spin correlations between the opposite edges.~\cite{Wakabayashi98,Costa11}
This behavior has a close analogy with spin excitations in $C$-type Heisenberg antiferromagnets. They are 
composed of one-dimensional chains with a FM superexchange embedded in a higher-dimensional system 
with AF couplings along the other directions. 
In such a system, the AF interactions dominate the actual form of spin-wave excitations: one finds linear 
Goldstone modes $\omega\simeq v_sq$  irrespective of the cubic direction while the FM interactions modify merely the 
spin-wave velocity  $v_s$.~\cite{Raczkowski02}

A comparison with Fig.~\ref{Skw}(b) reveals that the magnetic spectrum is directly affected by nonlocal interactions 
with spin-wave-like excitations continuing  to disperse up to larger wavevectors. 
We attribute this to the transfer of spectral weight in the single-particle spectrum at the edge
$A_e(q|{\pmb a}_1|=\pi,\omega)$  in Fig.~\ref{gap}(a) which reduces Landau damping at low frequencies. 
In contrast, despite the interaction-induced renormalization of the electron velocity at the Dirac points,  
which should in turn increase the spin-wave velocity, the spin spectral function remains essentially identical 
for small momenta $q|{\pmb a}_1|\le\frac{4\pi}{15}$, see the inset in Fig.~\ref{Skw}(b). 
This can be understood as resulting from the simultaneous reduction of the effective on-site repulsion by nonlocal 
interactions (or equivalently the reduction of the magnetic moment at the edge $m_{L/2}$ in Fig.~\ref{edge2}(a) by 
enhanced charge fluctuations) such that both renormalization effects effectively cancel each other out.

Figure~\ref{Skw_b} displays the dynamical spin structure factor along the bulk sites $S_b(q,\omega)$.
In contrast to the edge spin spectrum featuring a low-energy spin-wave-like mode, 
the bulk spin spectral function is broad and corresponds to the particle-hole continuum. 
In addition to the low-energy spectral weight in the long-wave limit, the spin excitations seem to also soften 
on approaching $q|{\pmb a}_1|=\tfrac{\pi}{2}$ and $q|{\pmb a}_1|=\tfrac{3\pi}{2}$ wavevectors. 
These two momenta correspond to wavevectors connecting two Dirac cones in the single-particle 
spectral function $A_b(q,\omega)$, see Fig.~\ref{Akw_b}. The softening at finite momenta  bears a striking similarity 
to the continuum of excitations with soft features around $q|{\pmb a}_1|=\tfrac{\pi}{2}$ 
and $q|{\pmb a}_1|=\tfrac{3\pi}{2}$ wavevectors  resolved in the dynamical charge structure factor
along the zigzag  edge $C_e(q,\omega)$, see Fig.~\ref{Ckw}(b).
The similarity between the low-energy spin and charge excitations at finite momenta confirms that they stems 
from scattering between the Dirac points.

\section{Summary}
\label{sec:summary}

By performing projective QMC simulations of a realistic model with long-range Coulomb interactions, 
we have revisited the stability of spin-polarized edge states of graphene nanoribbons with up to $N=480$ lattice sites. 
The tendency towards the extended FM spin polarization along the edges promoted by the on-site interaction remains 
\emph{dominant} over the competing  short-range charge correlations favored by long-range Coulomb interactions. 
Our results agree with the findings of Shi and Affleck using an effective edge model,~\cite{Affleck17} 
providing further evidence in support of edge magnetism for moderate values of nonlocal interactions.

While the reduction of spin correlations along the edge can be understood as resulting from the tendency 
of nonlocal Coulomb interactions to renormalize downwards the effective on-site interaction, the evolution 
of the single-particle spectrum cannot be accounted for within a simplified model with a purely local Hubbard 
interaction.  Indeed, on the basis of our QMC simulations we have found that spin and charge fluctuations 
are intrinsically mixed and both are of vital importance to the actual form of the dispersion relation of edge states. 
In particular,  increasing the relative  strength of nonlocal interactions  $\alpha$ with respect to the on-site repulsion 
tunes the single-particle gap at the $q|{\pmb a}_1|=\pi$ momentum. Thus, given that the screening length in graphene can be 
efficiently engineered by substrate modification,~\cite{Hwang12} deposition of a zigzag nanoribbon on substrates 
with a different dielectric constant should allow one to modify accordingly the low-energy electronic states.

Our final remark concerns doped zigzag nanoribbons. It has been proposed within the Hartee-Fock approximation of the 
Hubbard model that the extra charge will accommodate at the edges forming topological defects, i.e., domain walls, 
across which there is a change in the phase of the edge spins.~\cite{Brey17} 
Given the tendency for charge topological solitons in doped nanoribbons, the inclusion of the long-range part of the 
Coulomb interaction shall drive the extra charge apart, possibly affecting the charge concentration in the domain walls 
and thus the distance between solitons. A more quantitative discussion requires a separate analysis which constitutes an 
interesting subject for future.

\begin{acknowledgments}
This work was supported by the German Research Foundation (DFG) through SFB 1170 ToCoTronics.
The authors gratefully acknowledge the computing time granted by the John von Neumann Institute for Computing (NIC) 
and provided on the supercomputer JURECA~\cite{jureca} at J\"ulich Supercomputing Centre (JSC).
\end{acknowledgments}


%

\end{document}